\documentclass[preprint2,twocolumn,times,tighten]{aastex631}
\usepackage{graphicx}	% Including figure files
\usepackage{amsmath}	% Advanced maths commands
\usepackage{amssymb}	% Extra maths symbols
\let\tablenum\relax
\usepackage{siunitx}

\begin{document}

\title{Statistics of Gas Density, Velocity, and Magnetic Fields in Cool-Core Galaxy Clusters}

\email{yuehu@ias.edu; *NASA Hubble Fellow}

\author[0000-0002-8455-0805]{Yue Hu*}
\affiliation{Institute for Advanced Study, 1 Einstein Drive, Princeton, NJ 08540, USA }

\author{Alex Lazarian}
\affiliation{Department of Astronomy, University of Wisconsin-Madison, Madison, WI, 53706, USA}

\author{Gianfranco Brunetti}
\affiliation{INAF—Istituto di Radioastronomia di Bologna, Via Gobetti 101, I-40129 Bologna, Italy}

\author{John A. ZuHone}
\affiliation{Smithsonian Astrophysical Observatory, 60 Garden St., Cambridge, MA 02138, USA}


\begin{abstract}
Understanding turbulence within the Intracluster Medium (ICM) of galaxy clusters is pivotal for comprehending their evolution and dynamics. Employing 3D magnetohydrodynamic (MHD) simulations of galaxy cluster mergers, we examine the statistical properties of gas density, magnetic fields, and velocity, particularly emphasizing the central regions spanning 400 kpc. The simulations are designed to resemble massive cool-core clusters such as Perseus, while varying the initial plasma  $\beta$ values (100, 200, and 500). Our findings indicate that while the statistical histogram distributions of gas density and velocity appear similar across different $\beta$ scenarios, their spatial distributions and morphological patterns exhibit noticeable differences. Through the application of the second-order structure function, we identified a scaling relation in velocity fluctuations, characterized by a slope of 1/2 and predominantly dominated by solenoidal components. Furthermore, our analysis reveals a pronounced anisotropy in both velocity and magnetic field fluctuations, with more significant fluctuations along the direction perpendicular to the magnetic fields. This anisotropy is scale-dependent, becoming more pronounced at smaller scales, and exhibits a decreasing trend in scenarios where the magnetic field is relatively weak, particularly at $\beta=500$. This suggests that the anisotropic nature of these fluctuations is predominantly regulated by the magnetic fields. Additionally, we test the efficacy of the Synchrotron Intensity Gradient (SIG) method for tracing magnetic fields in these environments. The SIG shows a global agreement with the magnetic field across all three $\beta$ scenarios, confirming the SIG's insensitivity to the medium's magnetization level.
\end{abstract}




\section{Introduction} 
\label{sec:intro}
Turbulence is ubiquitous throughout the cosmos \citep{1995ApJ...443..209A, 2010ApJ...710..853C,2014Natur.515...85Z,2017ApJ...835....2X,2020ApJ...889L...1L,2020NatAs...4.1064H,2022ApJ...934....7H,2022ApJ...941...92H,2023arXiv230106709Z,2023MNRAS.524.2945G}. Among the most significant turbulent motions are those found within galaxy clusters \citep{2013MNRAS.432.3401G,2014Natur.515...85Z,2016Natur.535..117H,2020ApJ...889L...1L,2023MNRAS.524.2945G}, which are the largest gravitationally bound structures in the universe since the Big Bang \citep{2008SSRv..134...93F,2012A&ARv..20...54F,2019SSRv..215...16V,2022SciA....8.7623B}. Turbulence plays a crucial role in the evolution of these clusters. It can amplify seed magnetic fields through the turbulent dynamo process \citep{1967PhFl...10..859K, 1989MNRAS.241....1R, 2010PhRvE..82d6314E, 2012PhRvL.108c5002B, 2016ApJ...833..215X,2022ApJ...933..131S,2022ApJ...941..133H}. The resulting amplified magnetic fields, combining with turbulence, then facilitate the generation of synchrotron radiation \citep{1959AZh....36..422G,1969ARA&A...7..375G,1979rpa..book.....R, 2001MNRAS.320..365B,2014IJMPD..2330007B}. Additionally, turbulence contributes to maintaining energy equilibrium in the intracluster medium (ICM) via mechanisms such as turbulent heating \citep{2007ARA&A..45..117M, 2008SSRv..134...71R, 2014Natur.515...85Z,2023JPlPh..89d9017S}, regulating the transfer of thermal energy within the cluster. Turbulence also plays a vital role in accelerating and re-accelerating cosmic rays (CRs) within the ICM \citep{1949PhRv...75.1169F, 1996SSRv...75..279V, 1966ApJ...146..480J, 2002PhRvL..89B1102Y, 2004ApJ...614..757Y, 2010A&A...510A.101F,2014IJMPD..2330007B, 2019MNRAS.489.3905S, 2020ApJ...894...63X,HLX21a,2023JHEAp..40....1Z}, leading to the formation of diffuse structures like radio halos, radio relics, and mini halos \citep{2002ARA&A..40..319C,2008SSRv..134...93F,2010Sci...330..347V,2013ApJ...762...78Z,2019SSRv..215...16V,2021MNRAS.502.2518S, 2021ApJ...923..245D, 2021MNRAS.504.3435B}. Despite its critical importance, our understanding of turbulence, especially statistics of velocity fluctuations, in the ICM is still developing, primarily due to the challenges inherent in assessing it through observations.

X-ray surface brightness images, obtained through Chandra and XMM-Newton observations, are widely used to investigate density fluctuations in the ICM down to scales of a few kpc. By employing a statistical linear relationship between density and velocity fluctuations, the typical amplitude of velocity fluctuations is found to be around 100 km/s on the scales of tens of kpc \citep{2006ApJ...640..691V, 2013MNRAS.432.3401G, 2014Natur.515...85Z, 2015MNRAS.450.4184Z, 2016Natur.535..117H,2023MNRAS.520.5157Z}. The correspondingly derived velocity spectra exhibit a slope that is closely aligned with the Kolmogorov turbulence \citep{2015MNRAS.450.4184Z}. Furthermore, velocity fluctuations in the centers of cool-core clusters can be probed using H$\alpha$ and CO emissions emanating from their filamentary structures \citep{2001MNRAS.328..762E, 2001AJ....122.2281C, 2014ApJ...785...44M, 2019A&A...631A..22O,2022MNRAS.510.2327M}. By analyzing these filaments' line-of-sight (LOS) velocity distribution on the plane-of-the-sky (POS), \cite{2020ApJ...889L...1L} and \cite{2023FrASS..1038613G} observed that, on scales smaller than approximately 10 kpc, the velocity fluctuations exhibit a characteristic slope steeper than the Kolmogorov's 1/3 scaling, aligning more closely with the scaling of 1/2, although the actual slope can vary for different clusters. These observational studies have largely advanced our knowledge of turbulence in the ICM.

In addition to direct observations, numerical simulations offer invaluable insights into the turbulence within the ICM \citep{2010ApJ...717..908Z, 2013ApJ...762...78Z, 2021MNRAS.500.5072M, 2022MNRAS.513..864B, 2023A&A...673A.131L}. For example, \cite{2013ApJ...762...78Z} explored the generation of turbulence and the consequent formation of mini-halos through MHD simulations of gas sloshing in cluster cores. \cite{2021MNRAS.500.5072M} conducted hydrodynamic simulations to study the anisotropy in stratified velocity fluctuations within the ICM. Cosmological simulations, such as those from the IllustrisTNG project \citep{2022MNRAS.513..864B, 2023A&A...673A.131L}, expand the scope of these studies, allowing for an examination of ICM turbulence in a broader context. Among these theoretical and numerical studies, \cite{2017ApJ...842...30L} made an important finding: in subsonic ICM turbulence, the gradient of turbulent fluctuations – including those in velocity, magnetic fields, and density – are perpendicular to the local magnetic field direction. Building on this fundamental property, \cite{2017ApJ...842...30L,2020ApJ...901..162H} introduced the Synchrotron Intensity Gradient (SIG) as a novel method for mapping magnetic fields within galaxy clusters.

The SIG method, initially tested with MHD simulations of turbulence \citep{2017ApJ...842...30L,ZhangLazarian2019,2019ApJ...887..258H,2022arXiv220806074H}, has been applied to observational data from the nearby cool-core cluster Perseus \citep{2020ApJ...901..162H} as well as to more disturbed clusters such as RXC J1314.4 -2515 and Abell 2345 \citep{2023arXiv230610011H}. The efficacy of SIG in tracing the ICM magnetic fields has been validated through comparison with polarization measurements in the radio relics of RXC J1314.4 -2515 and Abell 2345. This comparison confirmed the efficiency of SIG in mapping the magnetic fields in these environments. However, it is important to note that the SIG method has not been tested with simulations specifically tailored to galaxy clusters with non-turbulence physics processes (like sloshing motions) included. Such a numerical test is important because one of the notable advantages of SIG is its insensitivity to the Faraday depolarization effect caused by thermal electrons and turbulent magnetic fields along the LOS \citep{2023arXiv230610011H}. This allows for the unique mapping of magnetic fields in the full cluster, including radio halos, which is one of the main challenges faced by the next-generation radio facilities, such as the Square Kilometre Array (SKA; \citealt{2013A&A...554A.102G,2019MNRAS.490.4841L}).

In light of the importance of ICM turbulence and the SIG method, this work aims to (1) investigate the statistical properties of gas density, velocity, and magnetic fields in galaxy cluster mergers, and (2) test the efficacy of SIG in these cluster environments. To achieve these goals, we have utilized simulations from the Galaxy Cluster Merger Catalog\footnote{\href{http://gcmc.hub.yt}{http://gcmc.hub.yt}} \citep{2018ApJS..234....4Z}, specifically a set of simulations originally presented in \cite{2011ApJ...743...16Z}. From this set, simulations were selected that have been evolved to a time scale where sloshing motions and turbulence have developed. This selection was made to closely approximate conditions in cool-core clusters such as the Perseus cluster \citep{2014Natur.515...85Z, 2015MNRAS.450.4184Z}, where SIG has been applied but lacks a corresponding polarization comparison. 

This paper is structured as follows: in \S~\ref{sec:data}, we detail the galaxy cluster simulations and the mock synchrotron observations utilized in our study. \S~\ref{sec:theory} briefly revisits the fundamental concepts of MHD turbulence anisotropy and the structure-function analysis for studying turbulence, alongside the SIG pipeline. In \S~\ref{sec:result}, we present our findings on turbulence characteristics, including velocity fluctuations, magnetic field fluctuations, and gas density fluctuations, in different plasma $\beta$ conditions and assess the effectiveness of SIG in tracing magnetic fields. \S~\ref{sec:dis} delves into the implications of our findings for understanding MHD turbulence in ICM and the potential applications of SIG. Finally, our conclusions are summarized in \S~\ref{sec:con}.

\section{Theoretical consideration: anisotropic MHD turbulence}
\label{sec:theory}
\subsection{Super-Alfv\'enic case}
MHD turbulence exhibits distinct characteristics depending on the relative strength of the magnetic field, typically quantified by the Alfv\'enic Mach number $M_A$. This number is the ratio of the turbulent injection velocity to the Alfv\'en velocity. In super-Alfv\'enic conditions ($M_A > 1$), where the magnetic field is dynamically weak, turbulence-induced magnetic field and velocity fluctuations were initially considered as isotropic above the Alfv\'en scale $l_A=L_{\rm inj}M_A^{-3}$ due to the diminished back-reaction of the magnetic field on the turbulence. Here $L_{\rm inj}$ is the injection scale of turbulence. 

However, as noted by \cite{2023arXiv230610011H}, advective flows, such as those induced by cluster mergers, can still regulate the dynamically unimportant magnetic field. These flows can introduce anisotropy to the magnetic fields, causing the field to align or elongate along the flows. This anisotropy means, at the same scale, magnetic field fluctuations are more pronounced along the direction perpendicular to the magnetic field. The intrinsic relationship between synchrotron emission and magnetic fields ensures that the anisotropy is reflected in the intensity of synchrotron emission. Thus, when taking the gradient of the synchrotron intensity, it is dominated by the perpendicular component, thereby enabling the probing of the magnetic field orientation in super-Alfv\'enic conditions.

\subsection{Sub-Alfv\'enic case:} Although the ICM is typically super-Alfv\'enic at large scales, it is crucial to note that the kinetic energy in turbulent motions often follows a Kolmogorov cascade, characterized by $\delta v_l \sim l^{1/3}$, where $\delta v_l$ is the velocity fluctuation at scale $l$. This scaling suggests that the magnetic back-reaction on turbulence becomes increasingly significant at smaller scales. Consequently, at the Alfv\'en scale $l_{\rm A}$, the turbulent velocity equates to the Alfv\'en velocity, transitioning super-Alfv\'enic turbulence to sub-Alfv\'enic \citep{2006ApJ...645L..25L}. The value of $l_{\rm A}$ has been found to range from 1 kpc to 60 kpc \citep{2023arXiv230610011H}. At scales smaller than $l_{\rm A}$, MHD turbulence, encompassing both magnetic field and velocity fluctuations, becomes anisotropic. This anisotropy results from the growing influence of the magnetic field on the dynamics of turbulence as the scale decreases, leading to a more complex and varied behavior of turbulence in the ICM.

Pivotal ideas in understanding this anisotropy are theories of MHD turbulence \citet{GS95} and turbulent reconnection in \citet{LV99}, which center around the concept of ``critical balance'' in the global mean field reference frame and local magnetic field reference frame, respectively. The latter is more important for using the anisotropy to trace the local magnetic field, and we briefly review it here.

In the local magnetic field reference frame, the ``critical balance'' posits a balance between the cascading time of turbulence $\delta v_{l,\bot}/l_\bot$ and the wave periods $v_A/l_\parallel$. Here $l_\bot$ and $l_\parallel$ represent the perpendicular and parallel scales of eddies relative to the local magnetic field, respectively, with $v_A$ being the Alfv\'en speed. $\delta v_{l,\bot}$ is the turbulent velocity at scale $l$ in the direction perpendicular to the local magnetic field. This dominance of the perpendicular component in turbulence cascading arises from the fact that in the presence of fast turbulent reconnection, the motions of eddies with size $l_\bot$ perpendicular to the local direction of the magnetic field are not suppressed,  offering minimal resistance for turbulent cascading. Therefore, turbulent energy predominantly cascades in the perpendicular direction, and the velocity fluctuation obeys scaling relations, such as the Kolmogorov one in the strong turbulence regime:
\begin{equation}
\label{eq.v}
 \delta v_{l,\bot}= \delta v_{\rm inj}(\frac{l_\bot}{L_{\rm inj}})^{\frac{1}{3}}M_A^{1/3},
\end{equation}
where $v_{\rm inj}$ is turbulence injection velocity at the turbulence injection scale $L_{\rm inj}$. Combining with the $''$critical balance$''$ condition yields the anisotropy scaling \citep{LV99}:
\begin{equation}
\label{eq.lv99}
 l_\parallel= L_{\rm inj}(\frac{l_\bot}{L_{\rm inj}})^{\frac{2}{3}}M_A^{-4/3},
\end{equation}
which has been demonstrated by numerical simulations \citep{CV20, 2001ApJ...554.1175M, CL03, 2010ApJ...720..742K, HXL21} and in-situ measurements in the solar wind \citep{2016ApJ...816...15W, 2020FrASS...7...83M, 2021ApJ...915L...8D, 2023arXiv230512507Z}.

Eq.~\ref{eq.v} provides the scaling relation for velocity fluctuations. The corresponding relations for density and magnetic field fluctuations can be derived from the linearized continuity and induction equations in Alfv\'enic turbulence \citep{CL03}: 
\begin{align}
\delta \rho_l&= \delta v_l\frac{\rho_0}{v_A}\mathcal{F}^{-1}(|\hat{\pmb{k}}\cdot\hat{\pmb{\xi}}|),\\
\delta B_l&= \delta v_l\frac{B_0}{v_A}\mathcal{F}^{-1}(|\hat{\pmb{B}}_0\cdot\hat{\pmb{\xi}}|), 
\end{align}
 where $\rho_0$ and $B_0$ denote the mean density and mean magnetic field strength. $\hat{\pmb{k}}$ and $\hat{\pmb{\xi}}$ represent the unit wavevector and displacement vector, respectively. $\mathcal{F}^{-1}$ denotes an appropriate inverse Fourier transform. %The density and magnetic field fluctuations are still dominated by their perpendicular components.

Given the implications of Eq.~\ref{eq.lv99}, which suggest $l_\parallel \gg l_\bot$, the gradient of the velocity, magnetic field, or density is dominated by the component perpendicular to the local magnetic field:
\begin{align}
\label{eq.vg}
    \nabla \delta v_l\approx\frac{\delta v_{l,\bot}}{l_{\bot}} \hat{\pmb{l}}_{\bot}&\approx\frac{\delta v_{\rm inj}}{L_{\rm inj}}M_A^{1/3}(\frac{l_{\bot}}{L_{\rm inj}})^{-2/3}\hat{\pmb{l}}_{\bot},  \\
    \nabla \delta \rho_l\approx\frac{\delta \rho_{l,\bot}}{l_{\bot}}\hat{\pmb{l}}_{\bot}&\approx\frac{\rho_0}{L_{\rm inj}}M_A^{4/3}(\frac{l_{\bot}}{L_{\rm inj}})^{-2/3}\mathcal{F}^{-1}(|\hat{\pmb{k}}\cdot\hat{\pmb{\xi}}|)\hat{\pmb{l}}_{\bot},  \\
        \nabla \delta B_l\approx\frac{\delta B_{l,\bot}}{l_{\bot}}\hat{\pmb{l}}_{\bot}&\approx\frac{B_0}{L_{\rm inj}}M_A^{4/3}(\frac{l_{\bot}}{L_{\rm inj}})^{-2/3}\mathcal{F}^{-1}(|\hat{\pmb{B}}_0\cdot\hat{\pmb{\xi}}|)\hat{\pmb{l}}_{\bot}.
\end{align}
The scaling relation for gradient is particularly important as it underscores that the gradient amplitude is inversely proportional to $l_\bot^{-2/3}$. This relation is derived under the assumption that the velocity fluctuations adhere to the Kolmogorov scaling. However, if these fluctuations follow the Burgers scaling, characterized by $ \delta v_{l,\bot} = (\frac{l_\bot}{L_{\rm inj}})^{1/2}\delta v_{\rm inj}$, the dependence of the gradient amplitude on the scale changes to  $l_\bot^{-1/2}$. In both scenarios, whether following Kolmogorov or Burgers scaling, the implication remains that the amplitude of turbulence-induced gradients increases at smaller scales. 
This property typically is not observed with non-turbulent-related gradients, such as velocity gradient induced by sloshing motion or differential rotation.

In both super-Alfv\'enic and sub-Alfv\'enic conditions, the density and magnetic fields fluctuations exhibit anisotropy. However, the underlying physical reasons for this anisotropy differ \footnote{This property alleviates constraints on the spatial resolution of observations since anisotropy occurs at both large-scale (i.e., poor resolution) and small-scale (i.e., high resolution) cases.}. As demonstrated in \citet{2020ApJ...901..162H} and \citet{2023arXiv230610011H}, synchrotron emission and X-ray residual maps (obtained by subtracting a beta model from the X-ray image) inherit the anisotropy present in magnetic fields and gas density. Consequently, their corresponding Synchrotron Intensity Gradient (SIG) and X-ray Intensity Gradient (XIG) are dominated by the component perpendicular to the magnetic field.

\section{Methodology}
\label{sec:data}
\subsection{Simulation setup}
The MHD simulations of idealized binary galaxy cluster mergers used in this study to produce sloshing and turbulent gas motions in a cluster core are from \citet{2011ApJ...743...16Z}. The simulations were generated from the parallel hydrodynamics/N-body astrophysical simulation code FLASH 3 \citep{2000ApJS..131..273F,2009arXiv0903.4875D}. The ideal MHD equations are solved using a directionally unsplit staggered mesh (USM) algorithm, which employs constrained transport, which guarantees that the evolved magnetic field satisfies the divergence-free condition \citep{1988ApJ...332..659E} to machine precision. The order of the USM algorithm corresponds to the Piecewise-Parabolic Method of \citet{1984JCoPh..54..174C}. The gravitational potential in the simulations is set up as the sum of two ``rigid bodies'' corresponding to the contributions to the potential from both clusters \citep{2012MNRAS.419.1338R}. 

In these simulations, a dark matter subcluster is set up to pass by the center of a ``main'' cool-core cluster similar to systems such as Perseus and Abell~2029. The radial profiles of gas temperature and density for the main cluster are set up in hydrostatic equilibrium in a dark matter-dominated gravitational potential well. A turbulent magnetic field is set up on the computational domain, which is initially set up using a Kolmogorov spectrum with the local average magnetic pressure being proportional to the ratio of the thermal and magnetic pressures via $\beta = p_{\rm th}/p_B$, which is different between different simulations. In this work, we examine the simulations with initial $\beta$ = 100, 200, and 500.

The gravitational interaction between the subcluster and the main cluster initiates ``sloshing'' gas motions in the main cluster's core. For this work, we used the state of each simulation after it had been evolved to an epoch of 3.15~Gyr, or roughly $\sim$1.6~Gyr after the core passage of the subcluster. For more details of the setup of the simulations, we refer the reader to Section 2 of \citet{2011ApJ...743...16Z}.

\subsection{Synthetic synchrotron observation}
To create a synthetic spectroscopic cube from our simulations, we utilize the density field $\rho(\pmb{x})$ and the magnetic field $\pmb{B}(\pmb{x})$, where $\pmb{x}=(x,y,z)$ represents spatial coordinates. The calculation of  the synchrotron intensity $I(x,y)$, Stokes parameters $Q(x,y)$ and $U(x,y)$, and polarization angle $\psi(x,y)$ maps is based on the following equations \citep{1959AZh....36..422G,1969ARA&A...7..375G,1990SvAL...16..297L}:
\begin{equation}
\begin{aligned}
\label{eq.ip}
I &\propto \int n_{e}(B_x^2 + B_y^2)B_{\bot}^{\gamma} dz,\\
Q &\propto -\int n_{e} (B_x^2 - B_y^2)B_{\bot}^{\gamma} dz,\\
U &\propto -\int n_{e}(2B_xB_y)B_{\bot}^{\gamma} dz,\\
\psi &= \frac{1}{2}\tan^{-1}(\frac{U}{Q}),
\end{aligned}
\end{equation}
where $B_{\bot}=\sqrt{B_x^2 + B_y^2}$ is the magnetic field component perpendicular to the LOS, with $B_x$ and $B_y$ being the $x$ and $y$ components, respectively. The term $n_{e}=\rho(\pmb{x})$ denotes the density of relativistic electrons, under the assumption that the density of relativistic particles is proportional to that of thermal particles. In practice, however, electrons in different regions may undergo distinct acceleration and cooling processes, potentially leading to variations in this proportionality and the observed synchrotron emission. Nevertheless, we expect that this assumption does not affect our conclusions regarding the SIG. This is because the SIG depends on the synchrotron intensity's anisotropy, which is fundamentally contributed by the anisotropy of the magnetic field itself. Although the distribution of relativistic particles is non-trivial, it is generally expected to also correlate with the magnetic field, given the field's essential role in particle acceleration.

Given that the anisotropy of synchrotron emission is relatively insensitive to the spectral index of the electron energy distribution \citep{2012ApJ...747....5L,2019ApJ...886...63Z}, we adopt a homogeneous and isotropic distribution $N(E)dE=N_0E^{-p}dE$ with a spectral index 
$p=3$. Here $N(E)$ represents the electron number density per unit energy interval $dE$. $N_0$ is the pre-factor of the electron distribution. This choice yields an index of $\gamma = (p - 3)/4$ for the synchrotron emission calculations. To make a clean comparison of polarization and SIG, Faraday rotation is not included here. Other constant factors at a given $p$ are not explicitly detailed in Eq.~\ref{eq.ip}, as they do not alter the statistics of the synchrotron emission.

%\section{Methodology}
\subsection{The second-order structure function}
\subsubsection{Compressive and Solenoidal decomposition}
In our study, we utilize the second-order structure function, a widely accepted method for quantifying the statistical properties of turbulent flows. The structure function of velocity is defined as:
\begin{equation}
\label{eq.sf}
{\rm SF}_2^v(r)=\langle|\pmb{v}(\pmb{x}+\pmb{r})-\pmb{v}(\pmb{x})|^2\rangle_r,
\end{equation}
where $\pmb{v}(\pmb{x})$ is the velocity at position $\pmb{x}$, and $\pmb{r}$ represents the separation vector. This function can be adapted to measure magnetic field or gas density fluctuations by replacing $\pmb{v}(\pmb{x})$ with $\pmb{B}(\pmb{x})$ or $\rho(\pmb{x})$, respectively. In the following, we will add the superscript ``$v$'', ``$B$'', or ``$\rho$'' to ${\rm SF_2}$ for distinguishing the variables accordingly.

Moreover, the fraction of solenoidal turbulence is an important aspect of the ICM study. We decompose the structure function into longitudinal (i.e., compressive) component ${\rm SF}_2^{v, {\rm com.}}(r)$ and transverse (i.e., solenoidal) component ${\rm SF}_2^{v, {\rm sol.}}(r)$, following the method used in \cite{2022MNRAS.513.2100H}. The decomposition is performed as:
\begin{align}
        \delta \pmb{v} &= \pmb{v}(\pmb{x}+\pmb{r})-\pmb{v}(\pmb{x}),\\
        {\rm SF}_2^{v, {\rm com.}}(r)&=\langle\delta \pmb{v}^2(\hat{\delta \pmb{v}}\cdot\hat{\pmb{r}})\rangle_r,\\
        {\rm SF}_2^{v, {\rm sol.}}(r)&={\rm SF}_2^v(r)-{\rm SF}_2^{v, {\rm com.}}(r).
\end{align}
Accordingly, the fraction the solenoidal component is ${\rm SF}_2^{v, {\rm sol.}}/{\rm SF}^v_2$.

\subsubsection{Decomposition with respect to local magnetic fields}
To investigate the anisotropy in turbulence within our simulations, we decompose the structure-function into components parallel and perpendicular to the local magnetic fields, following the methodology described in \cite{CV20}:
\begin{equation}
\label{eq.sf_loc}
\begin{aligned}
&\pmb{B}=\frac{1}{2}(\boldsymbol{B}(\pmb{x}+\pmb{r})+\pmb{B}(\pmb{x})),\\
&{\rm SF}^v_2(R,z)=\langle|\pmb{v}(\pmb{x}+\pmb{r})-\pmb{v}(\pmb{x})|^2\rangle_r,
\end{aligned}
\end{equation}
where $\pmb{B}$ defines the local magnetic field direction in a cylindrical coordinate system, with $z$-axis parallel to $\pmb{B}$, $R=|\hat{z}\times\pmb{r}|$ and $z=\hat{z}\cdot\pmb{r}$, where $\hat{z}=\pmb{B}/|\pmb{B}|$.

From these, we can derive the parallel ($\delta v_\parallel$) or perpendicular ($\delta v_\bot$) velocity fluctuations relative to the local magnetic fields as follows:
\begin{equation}
\label{eq.sf_v}
\begin{aligned}
\delta v_{\bot}(r)&=\sqrt{{\rm SF}_2^v(R,0)},\\
\delta v_{\parallel}(r)&=\sqrt{{\rm SF}_2^v(0,z)}.
\end{aligned}
\end{equation}

Similarly, the parallel ($\delta B_\parallel$) or perpendicular ($\delta B_\bot$) magnetic field fluctuations are obtained from:
\begin{equation}
\label{eq.sf_b}
\begin{aligned}
\delta B_{\bot}(r)&=\sqrt{{\rm SF}_2^B(R,0)},\\
\delta B_{\parallel}(r)&=\sqrt{{\rm SF}_2^B(0,z)}.
\end{aligned}
\end{equation}
This approach enables a nuanced analysis of turbulence characteristics in relation to the magnetic field.

\subsection{SIG pipeline}
The SIG pipeline, as described in \cite{2023arXiv230610011H}, is summarized in the following steps:

\textbf{Step 1.} The synchrotron intensity map $I(x, y)$ is convolved with the 3$\times$3 Sobel kernels.
%$G_x$ and $G_y$:
%\begin{equation}
%    \begin{aligned}
%        \nabla_{x} I(x, y) = G_x * I(x, y), \\
%        \nabla_{y} I(x, y) = G_y * I(x, y),
%    \end{aligned}
%\end{equation}
%where the asterisks denote convolutions. $\nabla_{x} I(x, y)$ and $\nabla_{y} I(x, y)$ are the gradient components along the $x$ and $y$ axis, respectively. They are used to calculate the overall pixelized gradient map $\psi_g (x,y)$:
%\begin{equation}
%       \begin{aligned}
%&\psi_g (x,y)=\tan^{-1}(\frac{\nabla_{y} I(x, y)}{\nabla_{x}I(x, y)}).
%\end{aligned}
%\end{equation}
From here, only pixels with an intensity less than three times the RMS noise are kept for the following steps. 

\textbf{Step 2.} The resulting $\psi_g(x,y)$ is then processed with the sub-block averaging method (\citealt{2017ApJ...837L..24Y}). Within each 16$\times$16-pixels sub-block, the histogram of gradient orientation is fitted with a Gaussian distribution, and the gradient orientation corresponding to the Gaussian distribution's peak is then taken as the most probable orientation of the gradient for that sub-block.

%The pixelized map $\psi_g(x,y)$ is divided into rectangular sub-blocks with a size of 16$\times$16 pixels. Other sizes can be used, but 16×16 has been verified as sufficient in numerical studies \citep{LY18a,2021ApJ...911...37H}. For each sub-block:
%\begin{enumerate}
%    \item A histogram of gradient orientation is created and the histogram is fitted with a Gaussian distribution.
%    \item The gradient orientation corresponding to the Gaussian distribution's peak is then taken as the most probable orientation of the gradient for that sub-block.
%\end{enumerate}
\begin{figure*}
	\includegraphics[width=1.0\linewidth]{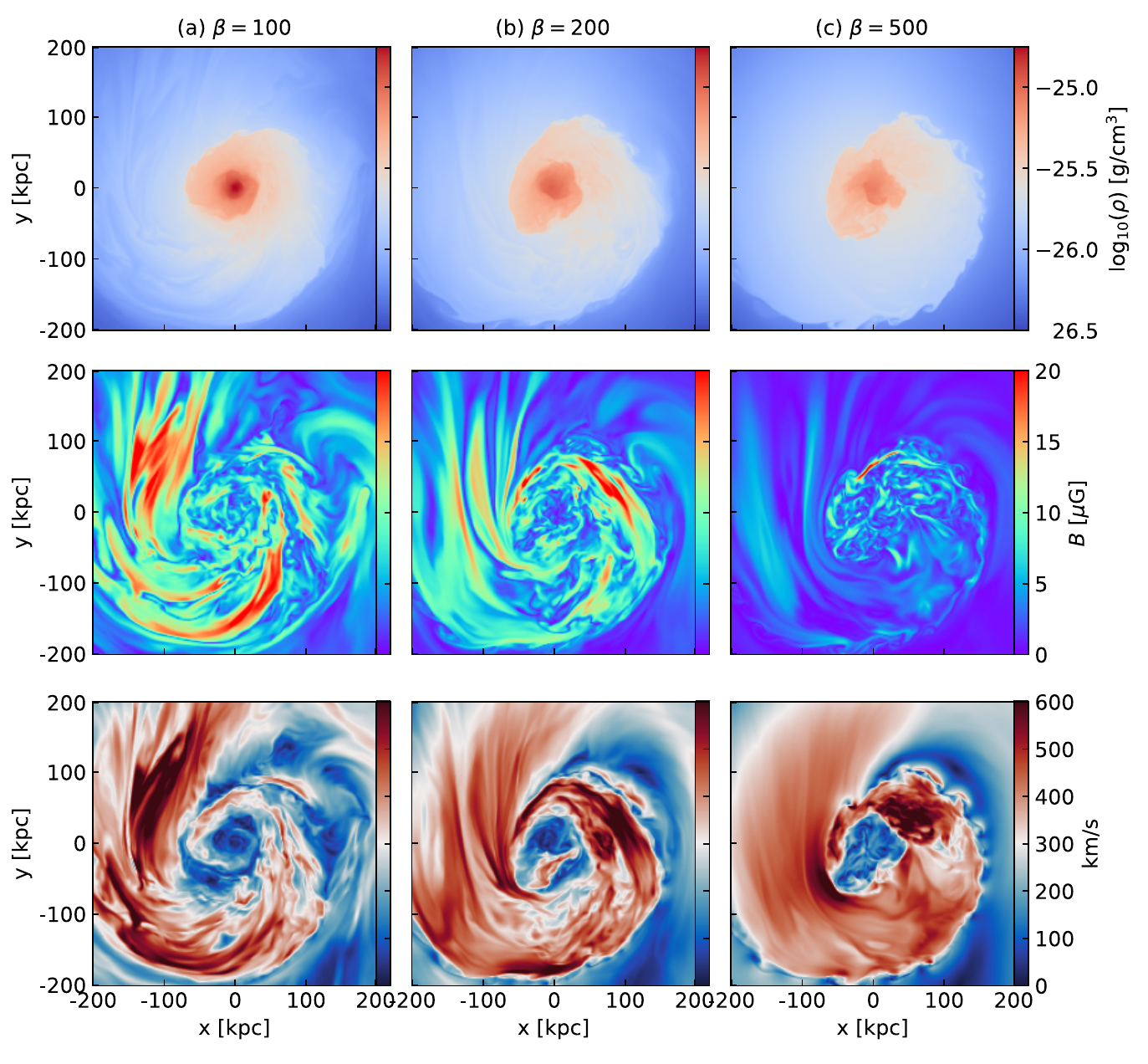}
    \caption{Maps of gas density (top), mean magnetic field strength (middle), and velocity (bottom) taken at the center $z=0$ kpc of the simulation box. The first row corresponds to the $\beta=100$ case, while the second row and third row represent $\beta=200$ and $\beta=500$, respectively.}
    \label{fig:2Dmap}
\end{figure*}

\begin{figure*}
	\includegraphics[width=1.0\linewidth]{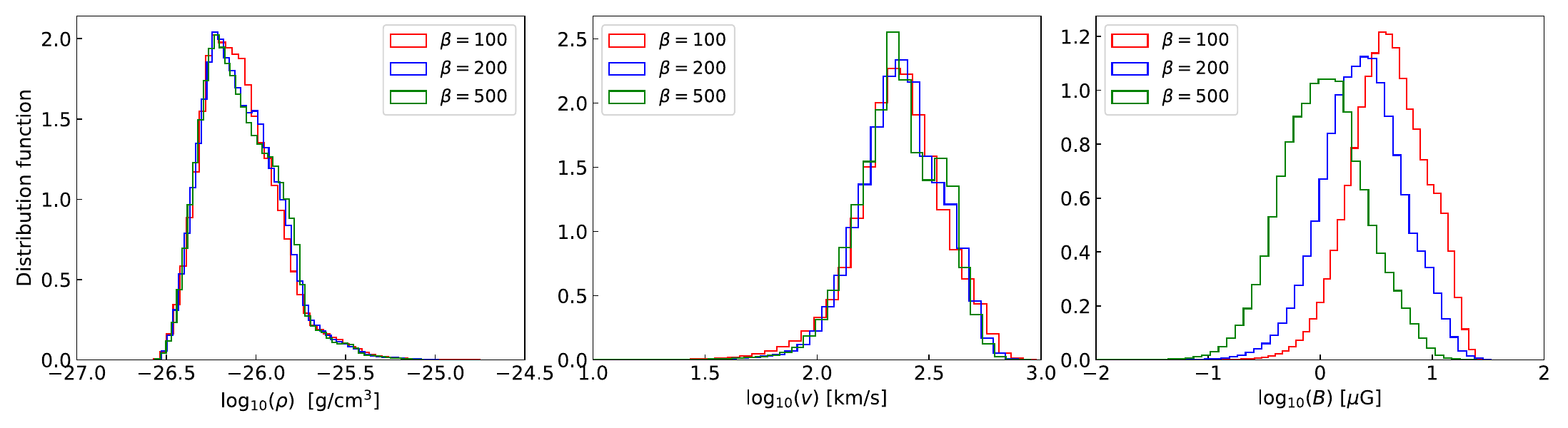}
    \caption{Histograms of gas density (left), velocity (middle), and magnetic field strength (right).}
    %anisotropy $l_\parallel/l_\bot$.}
    \label{fig:PDF}
\end{figure*}

\textbf{Step 3.} After the sub-block averaging, we obtain the averaged gradient angle map $\psi_{gs}(x,y)$ and construct the pseudo-Stokes parameters $Q_g(x,y)$ and $U_g(x,y)$ as follows:
%However, since the averaging procedure for each sub-block is independent, and actual magnetic field lines are correlated, we smooth the pseudo-Stokes parameters to correlate the averaged gradient with that of its neighboring. The pseudo-Stokes parameters $Q_g(x,y)$ and $U_g(x,y)$ are constructed as follows:
\begin{equation}
\label{eq.QU}
\begin{aligned}
    &Q_{\rm g} (x,y)  =  I(x, y) \cos(2\psi_{gs}(x,y)),\\
    &U_{\rm g} (x,y)  =  I(x, y) \sin(2\psi_{gs}(x,y)).
\end{aligned}
\end{equation}
%Weighted intensity ensures Gaussian distributions of  $Q_{\rm g}$ and $U_{\rm g}$, facilitating Gaussian filter smoothing with FWHM equal to the sub-block size. 
The POS magnetic field orientation can be inferred as:
\begin{equation}
\psi_{\rm B}(x,y) = \frac{1}{2}\tan^{-1}\left(\frac{U_{\rm g} (x,y)}{Q_{\rm g} (x,y)}\right) + \frac{\pi}{2}.
\end{equation}
The magnetic fields mapped by SIG in this case are intensity-weighted, similar to synchrotron polarization inferred magnetic fields.

To quantify the agreement between SIG and the magnetic field inferred from polarization, we utilize the Alignment Measure (AM; \citealt{GL17}), expressed as:
\begin{equation}
\begin{aligned}
{\rm AM} = 2 (\cos^2\theta_{\rm r} - \frac{1}{2}),
\end{aligned}
\end{equation}
where, $\theta_{\rm r} = \vert\phi_{\rm B}-\psi_{\rm B}\vert$. $\phi_{\rm B}=\psi+\pi/2$ is the magnetic field angle inferred from polarization. An AM value of 1 implies parallel alignment of $\phi_{\rm B}$ and $\psi_{\rm B}$, while -1 indicates perpendicularity.

\begin{figure}
\includegraphics[width=1.0\linewidth]{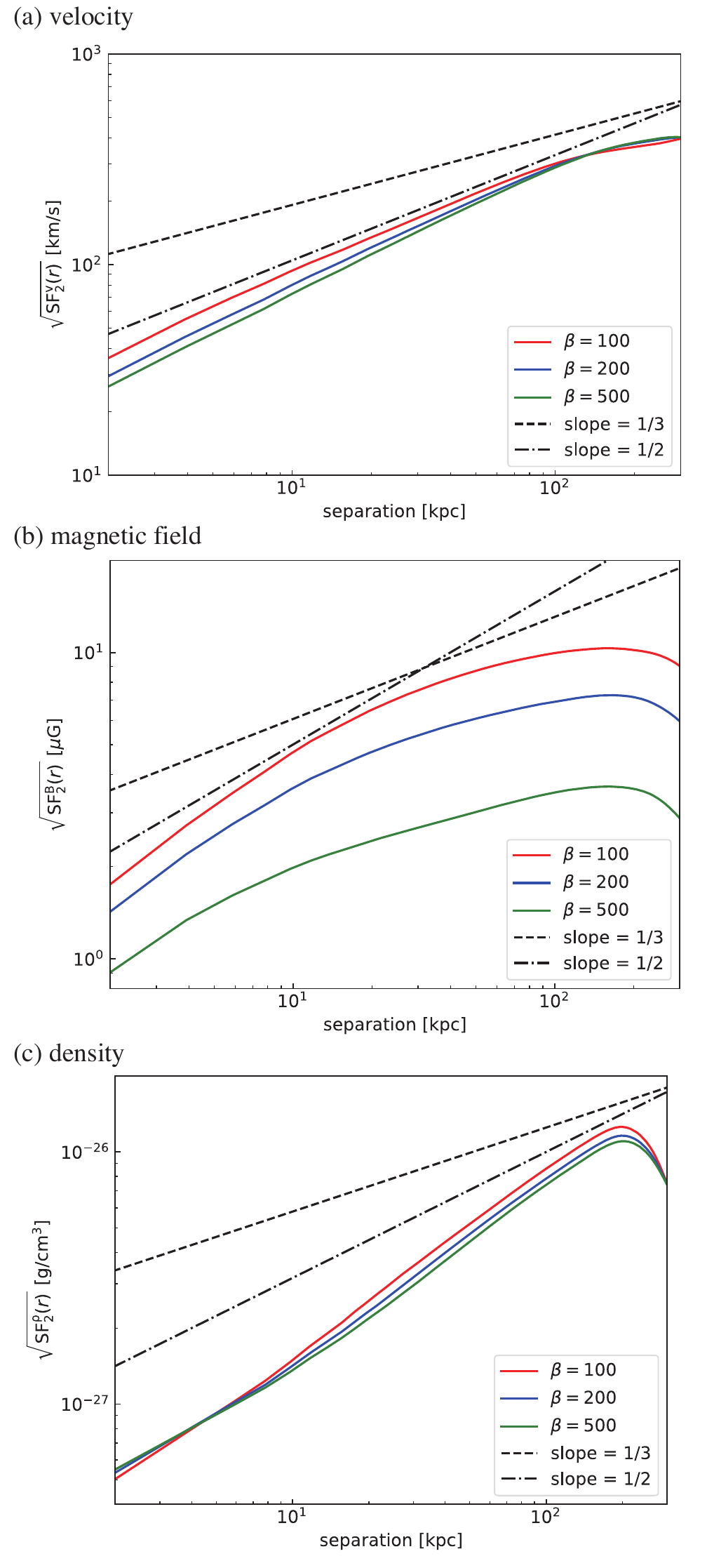}
        \caption{The square root of the second order structure function. The structure-function is calculated for gas velocity (top), magnetic field (middle), and gas density (bottom). To guide the eye, the dashed and dash-dotted lines represent power-law slopes of 1/3 and 1/2, for comparison with a Kolmogorov and Burgers scaling of turbulence, respectively.}
    \label{fig:SF_tot}
\end{figure}

\begin{figure}
\includegraphics[width=1.0\linewidth]{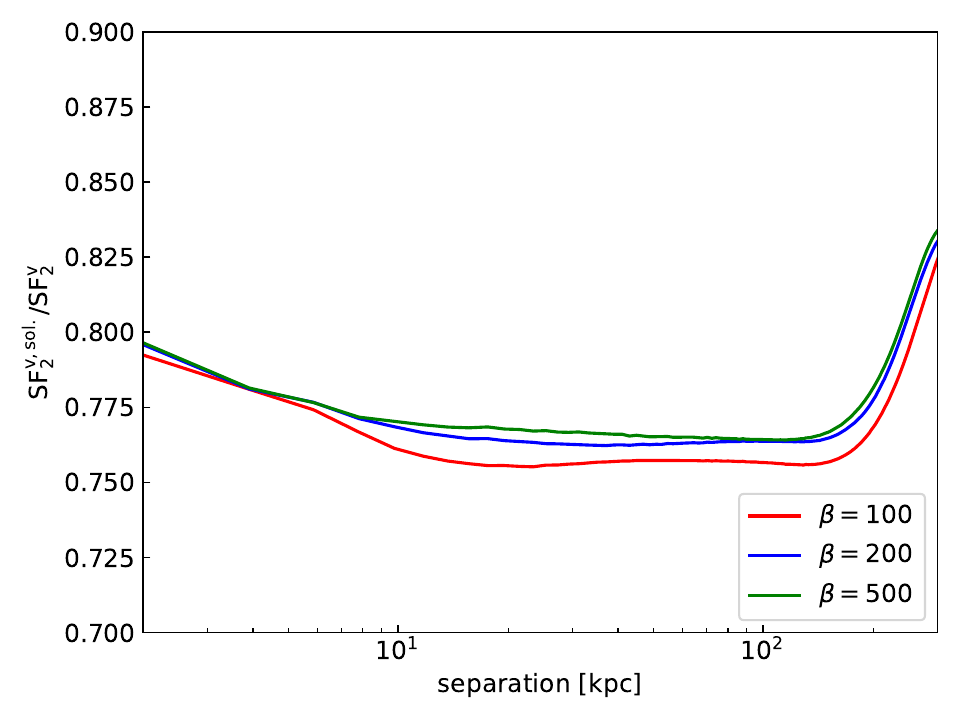}
        \caption{The plots present the fraction of the solenoidal (i.e., transverse) component of the velocity SF as a function of separation. }
    \label{fig:SF_sol}
\end{figure}

\section{Results}
\label{sec:result}
\subsection{Magnetic field can regulate the distribution of gas density and velocity}
Fig.~\ref{fig:2Dmap} presents the 2D slices of the gas density, magnetic field strength, and gas velocity. The slices are taken at the center $z=0$ kpc of the simulation box. In the gas density maps, a noticeable difference among the three cases ($\beta=100$, $\beta=100$, and $\beta=500$) emerges in terms of the concentration of density. Specifically, the case with $\beta=100$ exhibits a higher density value or a more concentrated density distribution in the center of the cluster compared to the $\beta=200$ and $500$ scenarios. The stronger magnetic field likely inhibits gas mixing of the cold and hot gas \citep{Hu2025}, which would tend to flatten out the density core in the center. On the other hand, a strong magnetic field could facilitate the removal of angular momentum, enhancing the accretion of gas. Despite this, the gas densities of these cases remain globally similar in magnitude. As shown in histograms of Fig.~\ref{fig:PDF}, the gas density ranges from $\approx3\times10^{-27}$ to $1\times10^{-25}$ g/cm$^3$, with a mean value around $\approx6\times10^{-27}$ g/cm$^3$. However, the density slice of the $\beta=500$ case reveals a more diffusive density distribution. This scenario is characterized by the emergence of more wave-like structures at the epicenter, which are indicative of the Kelvin–Helmholtz instability (KHI; \citealt{1961hhs..book.....C}). 

In our examination of the magnetic field across the three cases, we note that the field strength approximately ranges from $\approx\SI{0.1}{\micro G}$ to $\sim \SI{20}{\micro G}$, although they exhibit distinct median strengths (see Figs.~\ref{fig:2Dmap} and \ref{fig:PDF}). Specifically, in the case with $\beta=100$, the magnetic field strength has a median value of $\approx\SI{6}{\micro G}$. As $\beta$ increases, the median strength decreases, showing values of $\approx\SI{3}{\micro G}$ for $\beta=200$ and $\approx\SI{1}{\micro G}$ for $\beta=500$. This trend suggests that in the $\beta=500$ scenario, the magnetic field plays a less significant role, and the plasma fluid's motion is more akin to hydrodynamic behavior. In addition, in the 2D slice, we observe that the strongest magnetic fields typically manifest within the sloshing arms of the cluster rather than at the cluster center. 

\begin{figure*}
  \centering
  \gridline{
    \fig{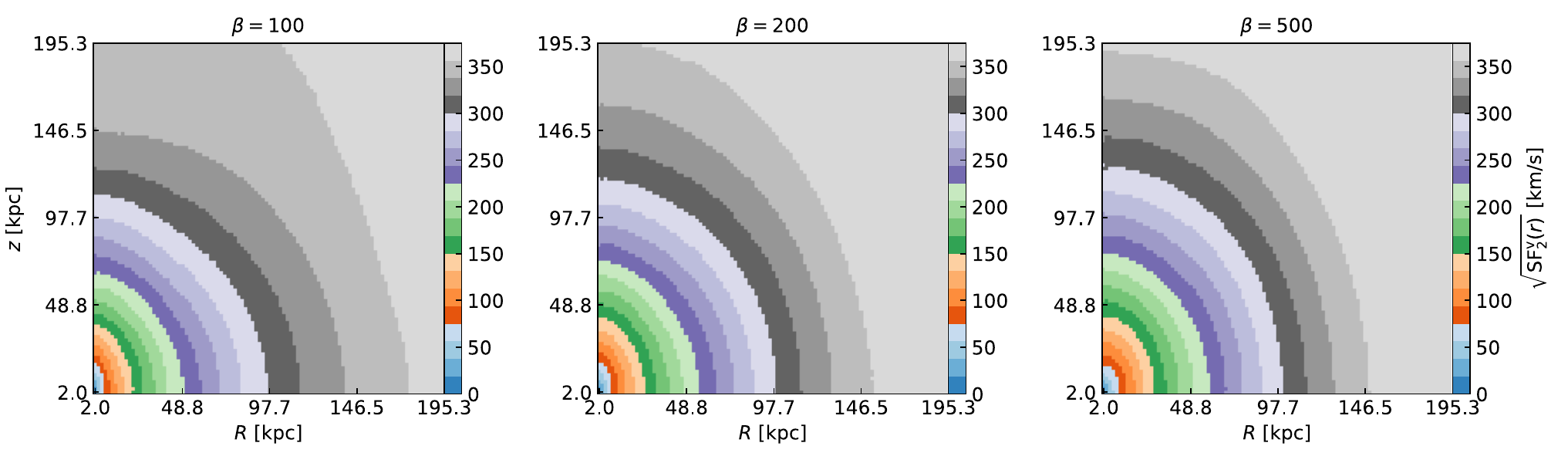}{0.99\linewidth}{(a) Velocity}
  }
  \vspace{-1.5em}
  \gridline{
    \fig{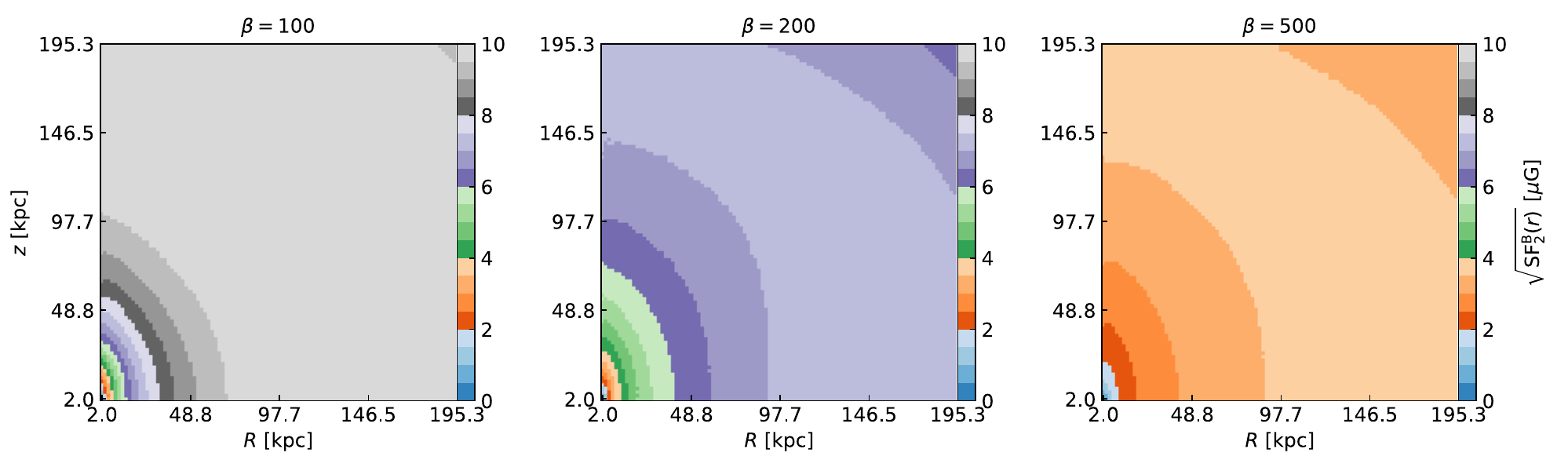}{0.99\linewidth}{(b) Magnetic field}
  }
  \caption{The map of velocity (panel a) and magnetic field (panel b) structure-functions decomposed along the directions parallel ($z$) and perpendicular ($R$) to the local magnetic field.}
    \label{fig:SF_decomposed}
\end{figure*}

Regarding velocity, the histogram (see Fig.~\ref{fig:PDF}) reveals distributions with a maximum velocity of approximately $700$~km/s and a median velocity of around $200$~km/s. However, the spatial distribution of velocity, as observed in the 2D slice (see Fig.~\ref{fig:2Dmap}), presents a different pattern. In the case with $\beta=100$, high velocities are typically found within the sloshing arms of the cluster, with the motion noticeably slowing down towards the center. The spatial co-occurrence of high velocity and strong magnetic fields in these regions suggests that the magnetic field could be amplified by the shearing of gas. As $\beta$ increases, the pattern shifts, and high-velocity motions tend to appear closer to the cluster center. This shift indicates a change in the dynamical interplay between the magnetic field and the gas motion, with the effects becoming more pronounced near the center of the cluster in higher 
$\beta$ scenarios.

\subsection{Magnetic field and velocity fluctuations are anisotropic}
\subsubsection{Total structure function}
Fig.~\ref{fig:SF_tot} presents the square root of the second-order structure function (as defined in Eq.~\ref{eq.sf}) for velocity, magnetic field, and gas density, effectively capturing the fluctuation levels of each respective quantity. Focusing on velocity, we observe a noticeable increase in fluctuation amplitude in the $\beta=100$ case. This increase is consistent with the spatial distribution of velocity seen in Fig.~\ref{fig:2Dmap}, where the velocity distribution in the low $\beta=100$ scenario appears less coherent compared to the higher $\beta$ cases. Given that the primary difference between these three cases is the magnetic field strength (as shown in Fig.~\ref{fig:PDF}), these stronger velocity fluctuations can be indicative of Magnetorotational Instability (MRI; \citealt{2015JPlPh..81e4908N}), which typically arises when a magnetic field destabilizes the differential rotation in ionized gas, leading to increased fluctuations.

\begin{figure*}
\includegraphics[width=1.0\linewidth]{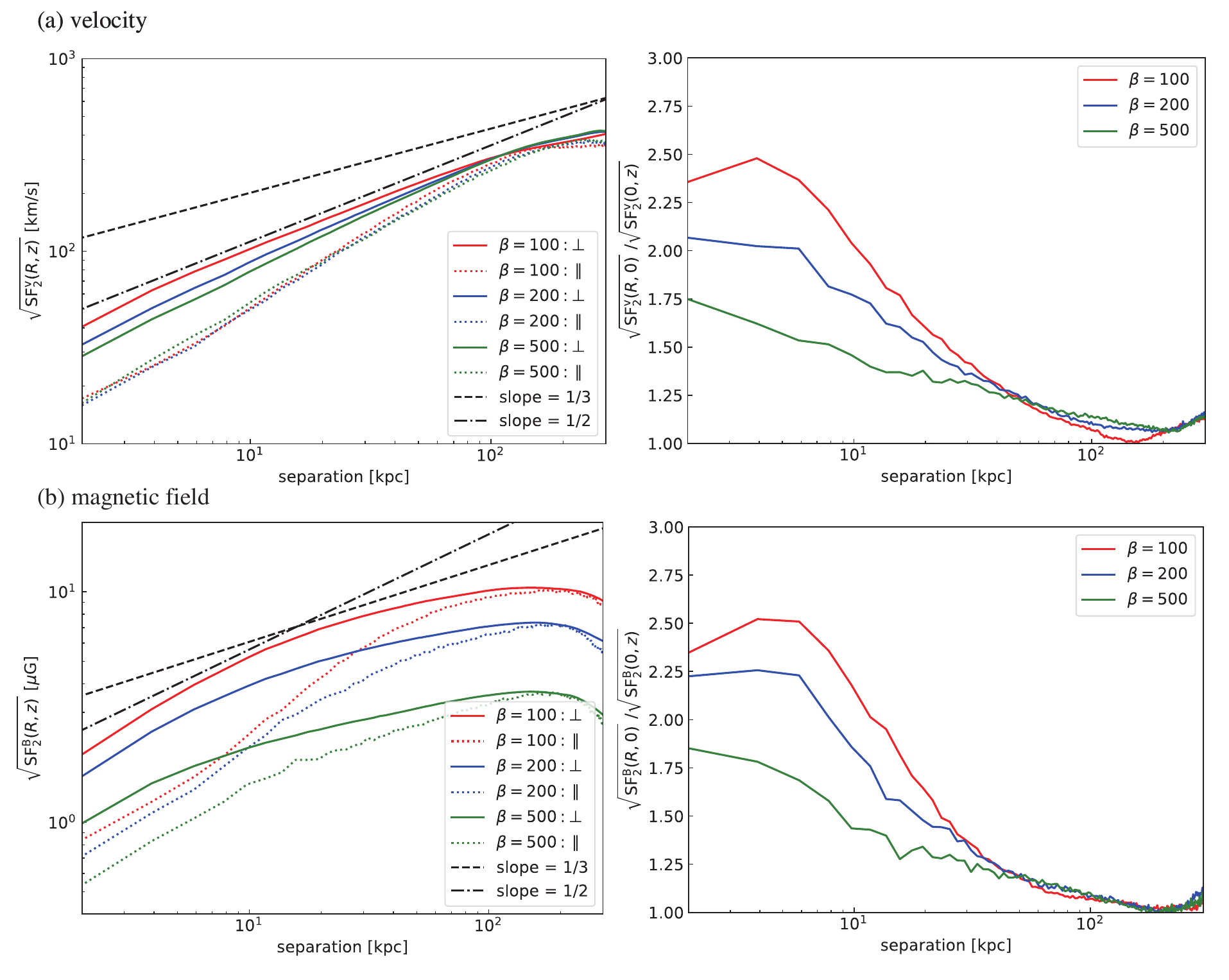}
        \caption{Left column: The square root of the decomposed structure function along the directions purely parallel ((0, $z$), dashed line) and perpendicular (($R$, 0), solid line) to the local magnetic field. The structure-function is calculated for gas velocity (panel a) and magnetic field (panel b). To guide the eye, the dashed and dash-dotted lines represent power-law slopes of 1/3 and 1/2, for comparison with a Kolmogorov and Burgers scaling of turbulence, respectively. Right column: The ratio of the decomposed structure functions, along the directions perpendicular ($R,0$) and parallel ($0,z$) to the local magnetic field, as a function of separation. The structure-function is calculated for gas velocity (panel a) and magnetic field (panel b).}
    \label{fig:SF_decomposed_plot}
\end{figure*}

Moreover, the structure functions of velocity closely align with the scaling predicted by Burgers turbulence, characterized by a slope of approximately $1/2$. This slope is steeper than the Kolmogorov scaling, which typically has a slope of $1/3$. The Burgers-like scaling is most evident in the low 
$\beta=100$ cases, holding up to scales (or separations) of about 10 - 100 kpc. Below this range, the structure functions exhibit a slightly steeper slope than the Burgers scaling. In parallel, the structure functions of the magnetic field also display a scaling that is steeper than the characteristic Burgers slope of $1/2$ at scales smaller than 10~kpc. Beyond this scale, the magnetic field structure functions have different trends from that of velocity fluctuations. Unlike the velocity case, the magnetic field structure functions become much shallower than a slope of $1/2$ at larger scales. Their slope, while closer, is still slightly shallower than the typical Kolmogorov slope of $1/3$.

In contrast to the velocity and magnetic field, the structure functions for gas mass density exhibit distinct scaling characteristics. These structure functions typically display a slope much steeper than $1/2$, indicative of less intense density variations at smaller scales. Additionally, regarding the amplitude of the structure-function, there is a notable increasing trend as $\beta$ decreases to 100. This increase in amplitude with lower $\beta$ values suggests stronger fluctuations in the low $\beta$ cases. Such a trend implies a more dynamic and variable density field in scenarios where the magnetic field strength is relatively strong ($\beta=100$), compared to cases with higher $\beta$ values.

\subsubsection{The energy fraction of the solenoidal component}
Fig.~\ref{fig:SF_sol} presents the energy fraction of the solenoidal (transverse) velocity component as a function of separation. The observed fractions, ranging between approximately 0.750 and 0.825, indicate a predominance of solenoidal velocity fluctuations. These fluctuations could be attributed to the differential rotation of the gas plasma around the cluster center. However, at scales smaller than $\sim10$~kpc, where the influence of differential rotation diminishes but numerical dissipation might start to become important, we notice an increase in the solenoidal fraction. Conversely, a decrease in the solenoidal fraction is observed with lower $\beta$ values. This trend suggests that in scenarios with stronger magnetic fields, the compressive component of velocity fluctuations grows somewhat faster than the solenoidal component, or alternatively, the magnetic fields may act to suppress the solenoidal component.

\begin{figure*}
  \centering
  \gridline{
    \fig{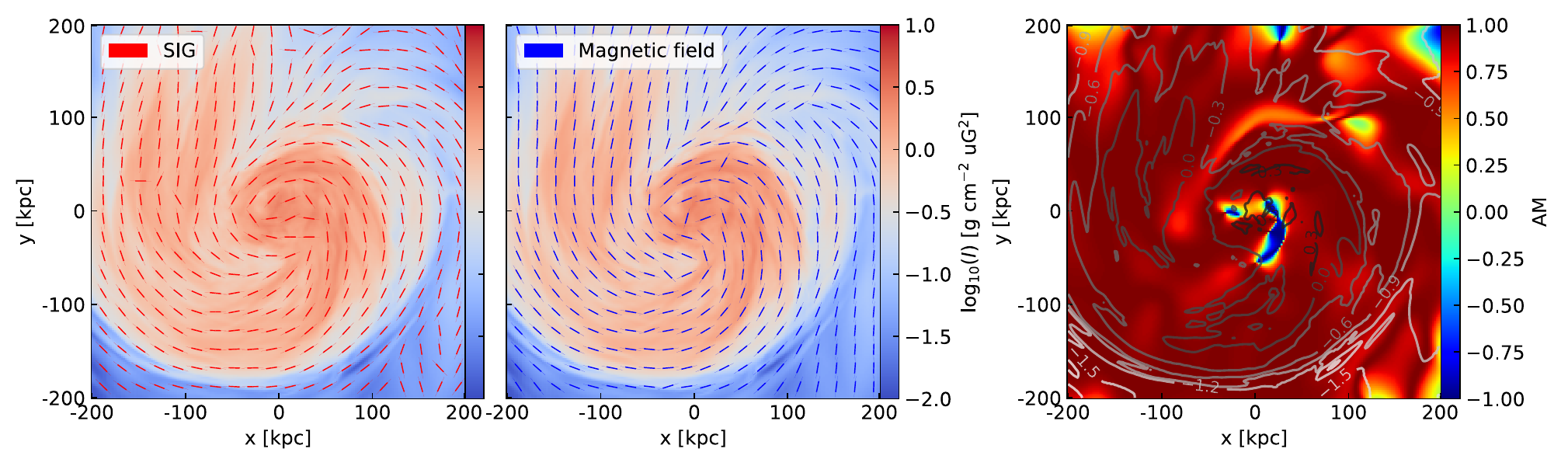}{0.99\linewidth}{(a) $x-y$ plane.}
  }
  \vspace{-1.5em}
  \gridline{
    \fig{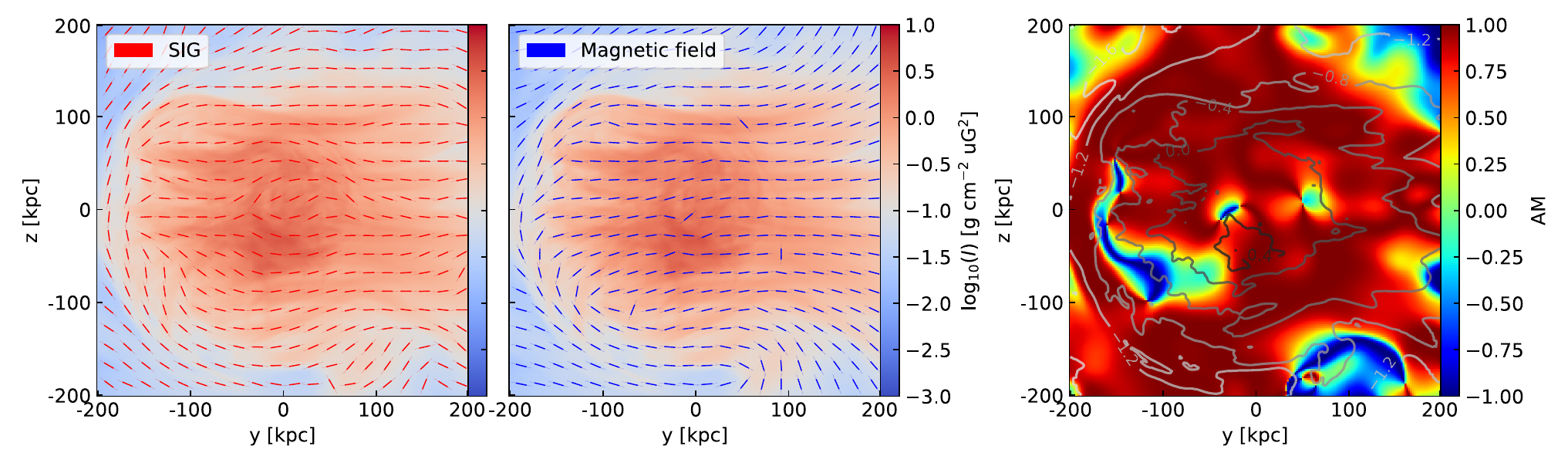}{0.99\linewidth}{(b) $y-z$ plane}
  }
  \caption{A comparison of the magnetic fields inferred from SIG (left, red segment) and polarization (middle, blue segment) for the $\beta=100$ case. The right panel shows the corresponding AM maps. The contours shown in the AM maps represent the logarithmic synchrotron intensity contours. Both face-on $x-y$ plane (panel a) and edge-on $y-z$ plane (panel b) cases are included.}
    \label{fig:sig_100}
\end{figure*}

\begin{figure*}
  \centering
  \gridline{
    \fig{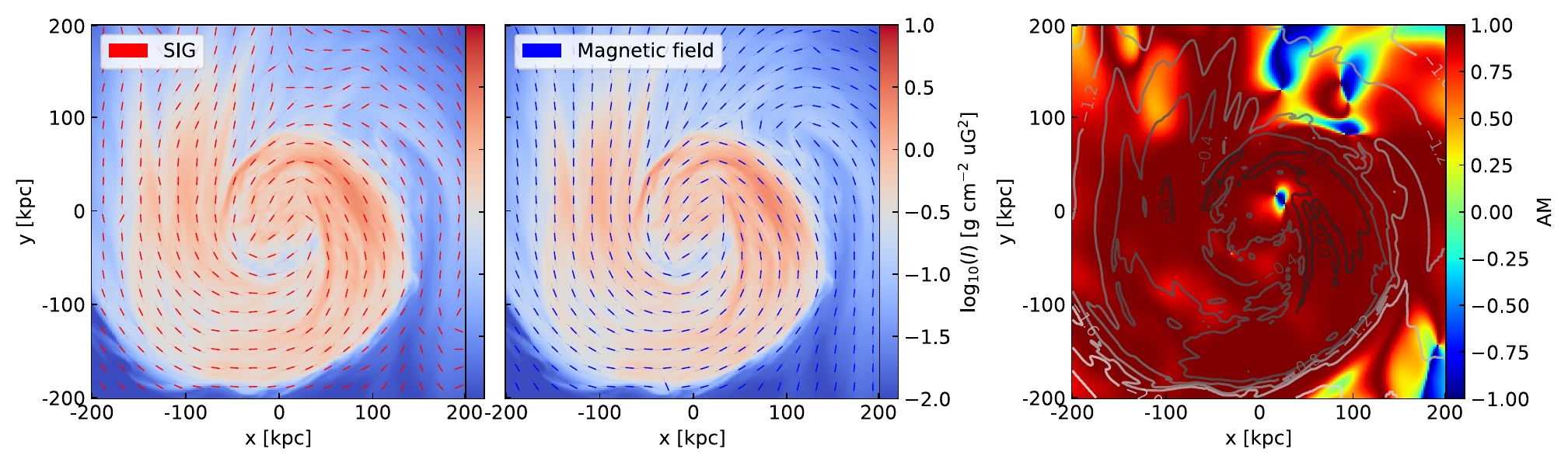}{0.99\linewidth}{(a) $x-y$ plane.}
  }
  \vspace{-1.5em}
  \gridline{
    \fig{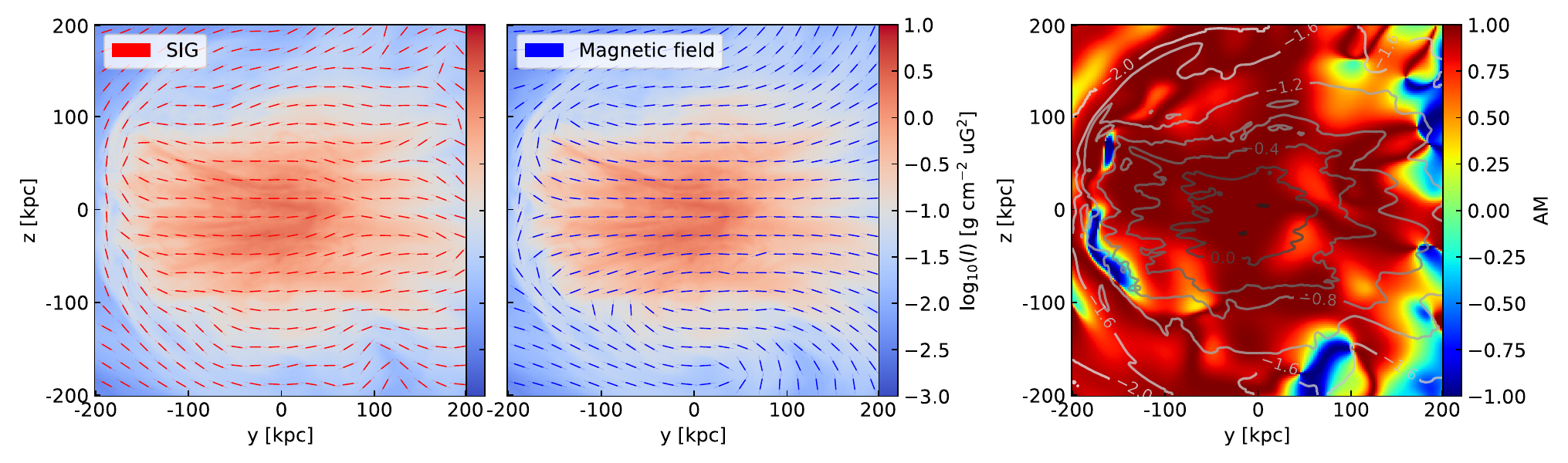}{0.99\linewidth}{(b) $y-z$ plane}
  }
    \caption{Same as Fig.~\ref{fig:sig_100}, but for the $\beta=200$ case.}
    \label{fig:sig_200}
\end{figure*}

\subsubsection{Decomposed structure function}
To further investigate the expected anisotropy inherent in MHD turbulence, we have undertaken a decomposition of the velocity structure functions (as outlined in Eq.~\ref{eq.sf_loc}) along axes parallel and perpendicular to the local magnetic fields. This result is presented in Fig.~\ref{fig:SF_decomposed}, where the $z$-axis corresponds to the direction parallel to the magnetic field, and the $R$-axis denotes the perpendicular direction. 

A distinct pattern is observed in the decomposed structure functions of velocity for the three cases ($\beta=100,200$, and 500). The contours of velocity fluctuations are elongated along the $z$-axis. This observation implies that, at any given separation scale (whether along the 
$z$-axis or the $R$-axis), the magnitude of velocity fluctuation is consistently greater in the direction perpendicular to the magnetic field (i.e., along the $R$-axis). This elongation is a clear indication of anisotropy in the velocity fluctuations, and by extension, the underlying MHD turbulence. A similar pattern of elongated contours is also observed in the case of magnetic field fluctuations, suggesting that the anisotropy occurs in magnetic fields as well.

Fig.~\ref{fig:SF_decomposed_plot} displays the square root of the decomposed structure function along directions purely parallel (along (0, $z$)) and perpendicular (along ($R$, 0)) to the local magnetic field. In the case of velocity fluctuations, it is observed that the perpendicular components generally exhibit higher amplitude at scales smaller than 100 kpc. Additionally, the overall statistical characteristics of the perpendicular component closely resemble the total structure function as seen in Fig.~\ref{fig:SF_tot}. Specifically, the scaling of the perpendicular component, especially in the $\beta=100$ case, aligns more closely with the scaling of 1/2. The slightly steeper scaling observed in the total structure function (Fig.~\ref{fig:SF_tot}) thus can be attributed to the steeper scaling of the parallel component. Regarding magnetic field fluctuations, the structure functions of the perpendicular components are highly consistent with the total structure function observed in Fig.~\ref{fig:SF_tot}. This consistency arises due to the predominance of perpendicular fluctuations. On the other hand, the structure functions of the parallel components exhibit a much steeper scaling slope than $1/2$, implying that parallel fluctuations decrease more rapidly at smaller scales.

In Fig.~\ref{fig:SF_decomposed_plot}, we conduct a quantitative analysis of the perpendicular and parallel velocity and magnetic field fluctuations. This analysis reveals a decreasing trend in the ratio of these fluctuations towards larger scales, indicating that the anisotropy is scale-dependent. This means fluctuations occurring perpendicular to the magnetic fields become relatively more significant at smaller scales. At scales of 2 - 5 kpc, perpendicular fluctuations (in both velocity and magnetic field) can exceed those parallel to the fields by a factor of approximately 2.5. However, at scales larger than 100 kpc, the ratio approaches 1, suggesting that the anisotropy in velocity fluctuations becomes less pronounced, leading to more isotropic statistics in both velocity and magnetic fields. 

Crucially, we note that the ratio of velocity and magnetic field fluctuations generally decreases with increasing $\beta$, which corresponds to weaker magnetic fields. This trend suggests that the anisotropic nature of these fluctuations is primarily caused by the magnetic fields. Specifically, in the case of $\beta=500$, the plasma fluid is less influenced by magnetic fields, tending towards isotropic hydrodynamic motion. This difference highlights the significant role that magnetic fields play in dictating the anisotropic properties of both velocity and magnetic field fluctuations in galaxy clusters.

This observed anisotropy, particularly the pronounced nature of fluctuations along directions perpendicular to the magnetic fields, provides essential validation for existing theoretical frameworks that synchrotron intensity gradients are predominantly perpendicular to magnetic fields. Consequently, the synchrotron intensity gradients could be used to trace the direction of magnetic fields.

\begin{figure*}
  \centering
  \gridline{
    \fig{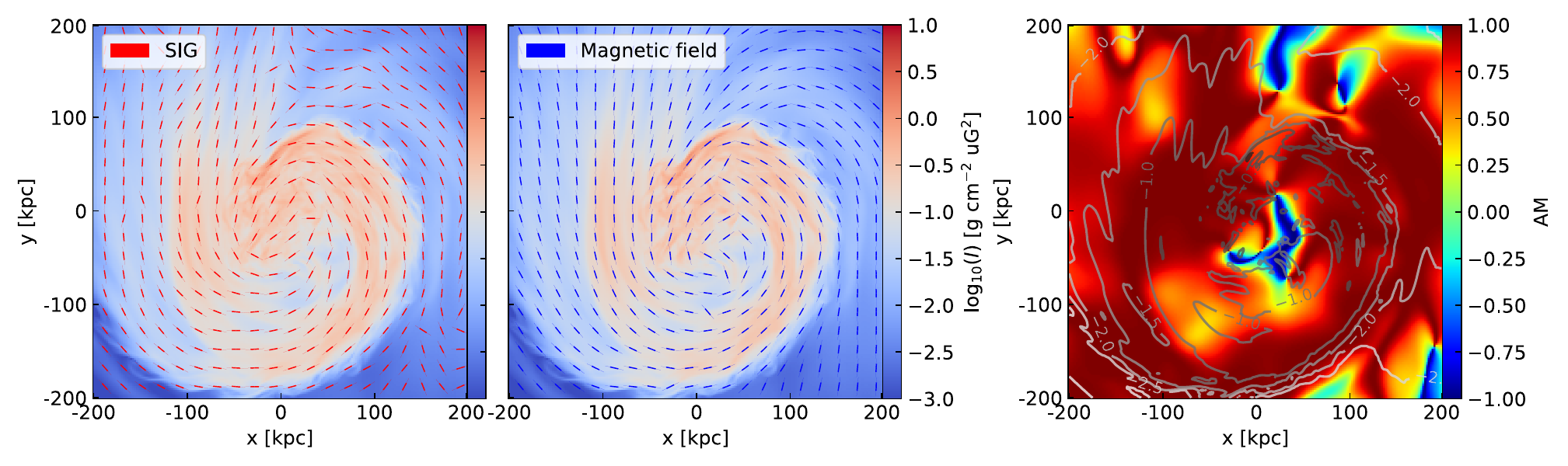}{0.99\linewidth}{(a) $x-y$ plane.}
  }
  \vspace{-1.5em}
  \gridline{
    \fig{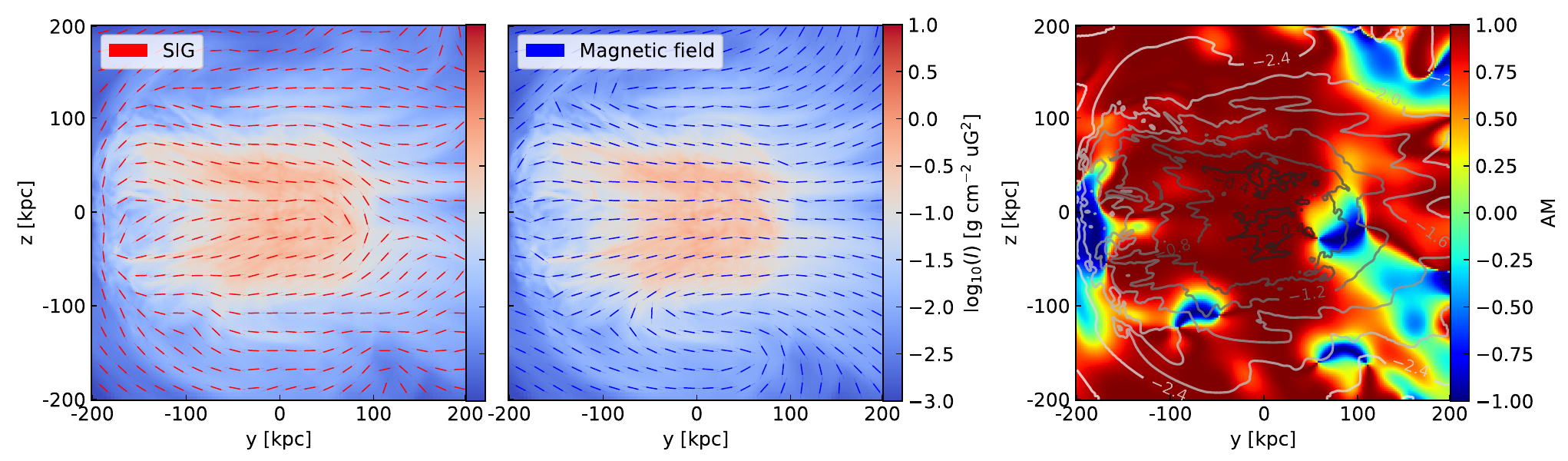}{0.99\linewidth}{(b) $y-z$ plane}
  }
    \caption{Same as Fig.~\ref{fig:sig_100}, but for the $\beta=500$ case.}
    \label{fig:sig_500}
\end{figure*}

\subsection{Comparison of SIG and polarization-derived magnetic fields}
In this section, we focus on testing the efficacy of SIG for tracing magnetic field orientations in different conditions of galaxy clusters. To ensure a comprehensive analysis, we rotate the orientation of our simulation boxes before generating the synthetic synchrotron emission. This allows us to mimic the varying observational angles that would be encountered in real observations, covering both face-on ($x-y$ plane) and edge-on ($y-z$ plane) orientations of clusters.

\textbf{The $\beta=100$ case:} Fig.~\ref{fig:sig_100} presents a comparative analysis of the magnetic field orientation in the $\beta=100$ environment, as derived from the SIG and inferred from polarization measurements. Additionally, the AM map, which quantifies the alignment between magnetic fields mapped by the SIG and those derived from polarization, is also presented. The SIG method utilizes the sub-block averaging method, which effectively reduces the resolution of the mapped magnetic fields. To ensure a fair comparison, the polarization-measured magnetic field maps were correspondingly smoothed to match the resolution of the SIG maps. This smoothing ensures that any observed differences are not caused by the difference in resolution.

In the face-on $x-y$ scenario, magnetic fields mapped by either the SIG or polarization show a spiraling pattern that is largely in sync with the synchrotron intensity structures of the clusters. The magnetic fields mapped by the SIG exhibit a general agreement with those deduced from polarization measurements. However, a misalignment between the SIG-mapped magnetic fields and those derived from polarization is observed in the cluster center. This misalignment leads to a negative AM distribution in this region.

In the edge-on $y-z$ scenario, we notice the magnetic fields still follow the synchrotron intensity structures, and there is a general agreement between the SIG-mapped magnetic fields and those inferred from polarization. However, some local misalignments are observed between the fields measured by the SIG and those derived from polarization. This is probably caused by the local variation in physical conditions and more discussion is given in \S~\ref{sec:dis}. The topology of the magnetic fields in this scenario reveals that the fields within the cluster are highly regulated by large-scale bulk flows along the merger axis. At the locations of cold fronts, the magnetic field is observed to be draped and amplified, leading to an orientation that is more perpendicular to the merger axis. This observation is consistent with the findings of \cite{2023arXiv230610011H}. Overall, the SIG method again demonstrates its general effectiveness in recovering the magnetic field orientation in galaxy clusters. 

\textbf{The $\beta=200$ and $\beta=500$ cases:} Figs.~\ref{fig:sig_200} and \ref{fig:sig_500} present the comparison of magnetic field orientations as mapped by the SIG and inferred from polarization for the $\beta=200$ and $\beta=500$ cases. Similar to the $\beta=100$ case, we observe that the SIG-derived magnetic fields generally align with those inferred from polarization methods. However, there are also instances of local misalignments, as indicated by negative values in the AM maps. These misalignments are typically found in three areas: the center of the clusters, the low-intensity peripheries, and the locations of cold fronts.

\subsection{Implication of the misalignment between the SIG and polarization-inferred magnetic fields}
\label{subsec:misalignment}
The observed misalignments between the SIG method and synchrotron polarization in mapping magnetic field orientations within galaxy clusters, including cases where the AM is positive but not 1, can be attributed to the difference in physical principles underlying each method. Synchrotron polarization emerges from the magnetic fields’ effects on relativistic electrons. On the other hand, the SIG method is based on the anisotropies in fluid flows to map the magnetic field orientation. This difference in underlying physical principles can lead to misalignment between the SIG method and polarization. Another plausible explanation for the observed misalignments in SIG measurement could be attributed to local variations in physical conditions. For instance, the cold fronts can provide extra gradients, which are not associated with fluid dynamics but may contribute to the total intensity gradient. In the case of perpendicular cold fronts, the rapid jump in intensity at the cold fronts creates an intensity gradient parallel to the magnetic field. This effect is the most likely reason for the misalignment observed at the sites of cold fronts.

\begin{figure}
    \centering
    \includegraphics[width=0.99\linewidth]{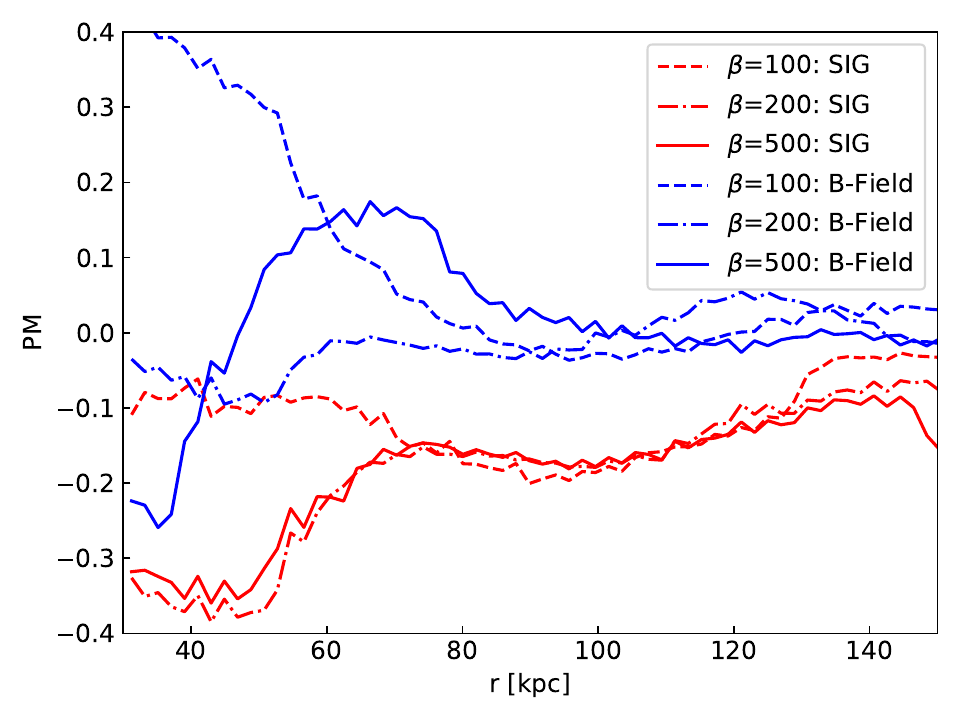}
    \caption{The relation of PM and the distance $r$ away from the cluster center. Negative PM implies the magnetic field tends to be tangential, while positive PM means the magnetic field follows the radial direction.}
    \label{fig:pm}
\end{figure}

\subsection{The magnetic field geometry in galaxy clusters}
Using the SIG method, \cite{2020ApJ...901..162H} found that in the Perseus cluster, the magnetic field is preferentially tangential near the cluster center, transitioning to a predominantly radial geometry at distances greater than approximately 40~kpc. To examine this transition, we use face-on cluster simulations. The magnetic field geometry is quantified using the Pitch Measure (PM; \citealt{2022ApJ...941..133H}), defined as:
\begin{equation}
\begin{aligned}
{\rm PM} = 2 (\cos^2\theta_{\rm p} - \frac{1}{2}),
\end{aligned}
\end{equation}
where $\theta_{\rm p}$ is the relative angle between the magnetic field, inferred from either SIG or polarization, and the radial direction. A negative PM indicates a tangential magnetic field geometry, while a positive PM corresponds to a radial magnetic field geometry.

Fig.~\ref{fig:pm} shows the relationship between PM and the distance from the cluster center. The magnetic field geometry exhibits distinct features depending on the plasma $\beta$ parameter. For a strong magnetic field case ($\beta=100$), the PM derived from polarization-based magnetic fields is positive (ranging from 0 to 0.4) within 80~kpc, indicating a preferentially radial magnetic field geometry. In contrast, the PM derived from SIG is negative (ranging from -0.2 to -0.1), suggesting a tangential field geometry. The radial field geometry implies that the magnetic field may facilitate angular momentum transfer, enabling gas to flow more easily toward the cluster center.

As $\beta$ increases, the PM for polarization-based magnetic fields decreases within 50~kpc and eventually becomes negative for $\beta=500$, indicating a transition to a tangential magnetic field geometry. Beyond 50~kpc, the PM transitions to positive values and eventually approaches zero at larger distances. For SIG-measured magnetic fields, the PM drops to around -0.4 for both $\beta=200$ and $\beta=500$ within 50~kpc, but it begins to increase beyond this distance. At distances between 50~kpc and 120~kpc, the PM reaches approximately -0.15 before eventually approaching zero. The transition scale of approximately 50~kpc is consistent with the findings of \cite{2020ApJ...901..162H}. Furthermore, the sensitivity of the transition scale to the value of $\beta$ suggests that the Perseus cluster likely has a plasma $\beta$ higher than 100.

\section{Discussion}
\label{sec:dis}
\subsection{Anisotropic velocity and magnetic field fluctuations}
This work thoroughly investigates the statistical properties of gas density, velocity, and magnetic fields in galaxy clusters, focusing on scenarios with initial plasma $\beta=100, 200$, and 500. Our results reveal that the slopes of velocity structure functions are steeper than those predicted by Kolmogorov turbulence (with a slope of 1/3), aligning more closely with 1/2. The velocity fluctuations are primarily dominated by the solenoidal component, with the fraction of the solenoidal fluctuations decreasing in scenarios where the magnetic field is relatively strong, as observed in the $\beta=100$ case. However, it should be noted that AGN feedback is not included in the simulations. With the presence of AGN feedback, the fraction of the compressive component may increase. 

%This increase in the fraction of the solenoidal component suggests that a strong magnetic field can induce more potential fluid motions along the magnetic field.

Furthermore, our analysis of decomposed velocity and magnetic field fluctuations, categorized into components parallel and perpendicular to local magnetic fields, demonstrates that perpendicular fluctuations are more significant and exhibit anisotropic behavior. The anisotropy is found to be more significant with decreasing scale and stronger magnetic fields, which corresponds to the theoretical expectation of the MHD turbulence \citep{LV99,CV20}, while other non-turbulence factors may also contribute. The differences in the spatial distribution of gas density and velocity across varying $\beta$ values further substantiate the profound impact of magnetic fields on the dynamics and evolution of galaxy clusters. In addition, the inclusion of AGN feedback is not expected to erase the anisotropy. In regions dominated by AGN-driven flows, dynamically weak magnetic fields are passively advected and tend to align with the flow direction. In turbulence-dominated regions, compressive motions consist of both slow and fast modes. The slow mode, like the Alfv\'en mode, is anisotropic, while the fast mode is isotropic \citep{CL03}. Therefore, unless the region is overwhelmingly dominated by fast modes, we can still expect the anisotropy.

\subsection{SIG: prospects and synergy with other methods}
The SIG method, as introduced in recent studies \citep{2017ApJ...842...30L}, offers an innovative way for mapping magnetic fields in galaxy clusters \citep{2020ApJ...901..162H,2023arXiv230610011H}. A fundamental premise of the SIG approach is that in super-Alfv\'enic conditions (scales larger than $l_A$) with weak magnetization, the magnetic fields passively align with large-scale flows induced by the cluster merger or AGN feedback. In such regimes, the gradients of synchrotron intensity, driven by magnetic field fluctuations, are preferentially oriented perpendicular to both the magnetic field and the local velocity field. This behavior was first demonstrated by \citet{2017ApJ...842...30L} and later confirmed in \citet{2023arXiv230610011H}. Conversely, under sub-Alfv\'enic conditions (i.e., at scales smaller than $l_A$), magnetic fields begin to dominate the dynamics, regulating the fluid motion. In this regime, both velocity and magnetic field fluctuations become anisotropic and predominantly perpendicular to the local magnetic field direction, as turbulent energy cascades more efficiently in directions perpendicular to the magnetic field \citep{2017ApJ...842...30L,2023arXiv230610011H}.

Our study confirms the applicability of the SIG method across both magnetization regimes. Specifically, we demonstrate that SIG successfully traces magnetic field orientations in weakly magnetized ($\beta = 500$) environments. This validation indicates that the SIG technique can be applied to radio observations with varying beam resolutions. When the beam size exceeds the Alfv\'en scale $l_A$, it corresponds to the super-Alfv\'enic regime; when smaller, it probes sub-Alfv\'enic motions. In both regimes, however, SIGs remain statistically perpendicular to the magnetic field, affirming their robustness across different magnetization conditions. Our results provide numerical support for the SIG-inferred magnetic field orientations in the Perseus cluster \citep{2020ApJ...901..162H} and in environments with relatively weak magnetic fields, such as the radio halos of RXC J1314.4-2515 and El Gordo \citep{2023arXiv230610011H}. The utility of SIG becomes especially compelling in light of recent LOFAR observations that have revealed diffuse synchrotron radiation on megaparsec scales, extending into cluster outskirts and intercluster bridges \citep{2019Sci...364..981G,2022hypa.confE...5B,2022Natur.609..911C}. With the advent of high signal-to-noise synchrotron intensity maps, SIGs are poised to trace magnetic field structures over vast cosmic volumes, offering new constraints on magnetogenesis theories and their role in shaping the large-scale structure of the universe.

A great value of the SIG technique lies in its demonstration that magnetic field information can be extracted from synchrotron intensity maps alone, supported by a theoretical foundation. Recognizing this opens the door to complementary approaches. Recent studies by \citet{2024ApJ...975...66H, 2024arXiv241107080Z} have shown that Convolutional Neural Networks (CNNs) can recover full 3D magnetic field information, including POS orientation, inclination with respect to the LOS, and strength, from radio observations. Applying CNNs to synchrotron observations of ICM holds significant promise for studying 3D magnetic fields in galaxy clusters. Moreover, the alignment between magnetic fields and cluster density structures, such as sloshing arms, supports the idea that X-ray intensity gradients, as suggested by \citet{2020ApJ...901..162H}, can also serve as tracers of magnetic field orientation.

\section{Summary}
\label{sec:con}
In this study, we have investigated turbulence within the ICM of galaxy clusters, utilizing 3D MHD simulations of galaxy cluster mergers with varying initial plasma $\beta$ of 100, 200, and 500. Our focus has been on analyzing the statistical properties of gas density, magnetic fields, and velocity, particularly in the central 400 kpc regions, and testing the efficacy of the Synchrotron Intensity Gradient (SIG) method in these environments. The major findings of our research are:
\begin{enumerate}
    \item Across different plasma $\beta$ scenarios, the statistical histogram distributions of gas density and velocity showed similarities. However, their spatial distributions and intensity features displayed noticeable variations, suggesting that magnetic fields significantly influence cluster dynamics and evolution.
    \item Velocity fluctuations were characterized by a power-law scaling slope close to 1/2. Solenoidal components largely dominated these fluctuations. The density fluctuations exhibited a characteristic slope steeper than 1/2. Magnetic field fluctuations show a variable slope: approximately 1/2 at scales smaller than 10 kpc but closer to 1/3 beyond this scale.
    \item Velocity and magnetic field fluctuations showed pronounced anisotropy, with greater amplitudes in the direction perpendicular to the magnetic fields. This anisotropy was scale-dependent, intensifying at smaller scales, and was less pronounced in scenarios with weaker magnetic fields ($\beta=500$).
    \item We tested the SIG method for tracing magnetic field orientations within these environments. The SIG results globally $\beta$ scenarios, demonstrating its independence from the ICM's magnetization level and its accuracy in mapping magnetic field orientations in galaxy clusters.
\end{enumerate}

%% Also note that the akcnowlodgment environment does not support long amounts of text. If you have a lot of people and institutions to acknowledge, do not use this command. Instead, create a new \section{Acknowledgments}.
\begin{acknowledgments}
Y.H. thanks Irina Zhuravleva for the helpful discussion. Y.H. acknowledges the support for this work provided by NASA through the NASA Hubble Fellowship grant No. HST-HF2-51557.001 awarded by the Space Telescope Science Institute, which is operated by the Association of Universities for Research in Astronomy, Incorporated, under NASA contract NAS5-26555. A.L. acknowledges the support of NSF grants AST 2307840. This work used SDSC Expanse CPU and NCSA Delta CPU through allocations PHY230032, PHY230033, PHY230091, PHY230105,  PHY230178, and PHY240183, from the Advanced Cyberinfrastructure Coordination Ecosystem: Services \& Support (ACCESS) program, which is supported by National Science Foundation grants \#2138259, \#2138286, \#2138307, \#2137603, and \#2138296. This work used data from the Galaxy Cluster Merger Catalog (http://gcmc.hub.yt).
\end{acknowledgments}


\vspace{5mm}

\software{Python3 \citep{10.5555/1593511}
          }


%\newpage
\bibliography{sample631}{}
\bibliographystyle{aasjournal}



\end{document}